\documentclass[10pt, conference]{IEEEtran}
\usepackage{array}
\usepackage{graphicx}
\usepackage{amsmath}
\usepackage{amssymb}
\usepackage{setspace}
\usepackage{algorithm}
\usepackage[]{algorithmic}
\usepackage{tikz}
\usetikzlibrary{shapes,arrows}
\usepackage{xcolor}
\usepackage{multirow}
\usepackage{paralist}
\usepackage{url}
\usepackage{etoolbox}
\usepackage{cancel}
\usepackage{tabu}
\usepackage{makecell}
\usepackage{mathtools}
\usepackage{marvosym}
\def\BibTeX{{\rm B\kern-.05em{\sc i\kern-.025em b}\kern-.08em
		T\kern-.1667em\lower.7ex\hbox{E}\kern-.125emX}}

\newcommand{\litmus}[0]{$\text{LITMUS}^{\text{RT}}$}

\floatname{algorithm}{Algorithm}
\newcommand{\algorithmicinput}{\textbf{Input:}}
\newcommand{\INPUT}{\item[\algorithmicinput]}

\begin{document}
	
	\title{Supporting Multiprocessor Resource Synchronization Protocols in RTEMS}
	
	\author{\IEEEauthorblockN{Junjie Shi\IEEEauthorrefmark{1}, Jan Duy Thien Pham\IEEEauthorrefmark{1}, Malte M\"unch\IEEEauthorrefmark{1}, Jan Viktor Hafemeister\IEEEauthorrefmark{1}, Jian-Jia Chen\IEEEauthorrefmark{1} and Kuan-Hsun Chen\IEEEauthorrefmark{2}\\}
		\IEEEauthorblockA{\IEEEauthorrefmark{1}Department of Computer Science, Technische Universitat Dortmund, Germany\\
		}
		\IEEEauthorblockA{ \IEEEauthorrefmark{2}Department of Electrical Engineering, Mathematics and Computer Science, University of Twente, the Netherlands\\
			E-mail: \{junjie.shi, jan.pham, malte.muench, jan.hafemeister, jian-jia.chen\}@tu-dortmund.de,
			k.h.chen@utwente.nl\\    
		}
	}	
	
	\pagestyle{plain}

	\maketitle
	
	\begin{abstract}
		When considering recurrent tasks in real-time systems, concurrent access to
		shared resources can cause race conditions or data corruption.
		Such a problem has been extensively studied since the 1990s, and numerous resource synchronization protocols have been developed for both uni-processor and multiprocessor real-time systems, with the assumption that
		the operating overheads are negligible.
		However, such overheads may also impact the performance of different protocols depending on the practical implementation, e.g., 
		resources are accessed locally or remotely,
		and tasks spin or suspend themselves when the requested resources are not available.
		In this paper, to show the applicability of different protocols in real-world systems, we detail the implementation of
		several state-of-the-art multiprocessor resource synchronization protocols in RTEMS. 
		To study the impact of the implementation overheads, we deploy these implemented protocols on a real platform with synthetic task sets. The measured results illustrate that the developed resource synchronization protocols in RTEMS are comparable to the officially supported protocol, i.e., MrsP.
	\end{abstract}
	
	\thispagestyle{plain}
	

	\section{Introduction}
	\label{sec:introduction}
	In multi-tasking real-time systems, 
	the accesses to shared resources, e.g., file, memory cell, etc., are
	mutually exclusive, to prevent race conditions or data corruptions.
	A code segment that a task accesses to the shared resource(s) is called a \emph{critical section}, which is protected by using binary semaphores or mutex locks.
	That is, 
	a task must finish its execution of the critical section before another task
	can access the same resource.
	However, the mutually exclusive executions of critical sections may cause other problems, i.e., 
	priority inversion and deadlock, which could jeopardize the predictability of the real-time system.
	In order to guarantee the timeliness of a real-time system, a lot of resource synchronization protocols have been developed and analyzed since 1990s for both
	uni-processor and multiprocessor real-time systems.
	
	In uni-processor real-time systems, the 
	Priority Inheritance Protocol (PIP) and the Priority Ceiling Protocol~(PCP) 
	by Sha~et~al.~\cite{DBLP:journals/tc/ShaRL90}, as well as the Stack
	Resource Policy~(SRP) by Baker~\cite{DBLP:journals/rts/Baker91}
	have been widely studied. 
	Since PIP may potentially lead to a deadlock requiring additional verification to avoid~\cite{DBLP:conf/icfem/GadiaAB16}, PCP has been relatively common and its performance has been widely accepted.
	Specifically, a variant of PCP has been implemented in Ada (named Ceiling locking) and in POSIX (named Priority Protect Protocol).
	
	Because of the 
	increasing demand of computational power of real-time systems,
	multiprocessor platforms have been widely used.
	A lot of multiprocessor resource synchronization protocols have
	been proposed and extensively studied in the domain, such as the Distributed 
	Priority Ceiling Protocol 
	(DPCP)~\cite{DBLP:conf/rtss/RajkumarSL88}, the
	Multiprocessor Priority Ceiling Protocol 
	(MPCP)~\cite{Rajkumar_1990},
	the Multiprocessor Stack Resource Policy
	(MSRP)~\cite{DBLP:conf/rtss/GaiLN01}, the Flexible Multiprocessor
	Locking Protocol (FMLP)~\cite{block-2007}, the $O(m)$ Locking Protocol
	(OMLP)~\cite{DBLP:conf/rtss/BrandenburgA10}, the Multiprocessor
	Bandwidth Inheritance (M-BWI)~\cite{DBLP:conf/ecrts/FaggioliLC10},
	gEDF-vpr~\cite{DBLP:journals/rts/AnderssonE10},
	LP-EE-vpr~\cite{DBLP:journals/rts/AnderssonR14},
	the Multiprocessor resource sharing Protocol
	(MrsP)~\cite{DBLP:conf/ecrts/BurnsW13},
	the Resource-Oriented Partitioned PCP (ROP-PCP)~\cite{RTSS2016-resource},
	the Dependency Graph Approach (DGA) for frame-based task set~\cite{Chen-Dependency-RTSS18},
	and its extension for periodic task set (HDGA)~\cite{Shi-Dependency-RTAS2019}.  
	
	Although the protocols above provide the timing guarantees by bounding the worst-case response time of tasks, most of them rely on the assumption that the overheads invoked by the implementation are negligible.
	However, rethinking of the assumption is in fact needed. Depending on their settings, e.g.,  local or remote execution of critical sections, multiprocessor scheduling paradigm, and the tasks' waiting semantics, the performance of different protocols is highly relevant to the implementation.
	For example, under a suspension-based synchronization protocol, tasks that are waiting for access  to a shared resource (i.e., the resource is locked by another task) are suspended. This strategy frees the processor so that it can be used by other ready tasks, which exploits the utilization of processor, but also increases the context switch overhead due to extra en-queue and de-queue operations for each suspension.
	In contract, under a spin-based synchronization protocol, the task does not 
	give up its privilege on the processor and has to wait by spinning on the processor until it can access the requested resource and starts its critical section, which is efficient when the critical sections are short~\cite{han-2014}. 	
	
	

	In fact, there are only a few of the protocols have been officially supported, and there are two real-time operating systems popular in the domain: the Linux Testbed for Multiprocessor Scheduling in Real-Time Systems
(\litmus)~\cite{calandrino2006litmus}, and 
Real-Time Executive for Multiprocessor Systems (RTEMS)~\cite{rtems}.
 	\litmus is an experimental platform for timing analysis mainly for academic usages. 
	Brandenburg et~al. implemented DPCP, MPCP, and FMLP~\cite{brandenburg2008implementation},
	Catellani et~al. implemented MrsP~\cite{DBLP:conf/adaEurope/CatellaniBHM15}, and Shi et~al. solidate the implementation of MrsP~\cite{shi-OSPERT17}. In addition, 
	the recently developed DGA and its extension for periodic tasks 
	HDGA have been implemented by Shi et~al. 
	in~\cite{DGALITMUS, HDGALITMUS}.
Alternatively, RTEMS is an open-source real-time operating system which is popular for industrial applications. RTEMS has been widely used in many fields, e.g., space flight, medical, networking, etc. However, in RTEMS, only MrsP implemented by Catellani et~al. in~\cite{DBLP:conf/adaEurope/CatellaniBHM15}, is officially supported in the upstream repository.

	Therefore, we believe it is beneficial to provide comprehensively support on RTEMS with resource synchronization protocols for the related researches. Afterwards, the performance of resource synchronization protocols might be clarified by system designers, and the optimizations of implementation can also be discussed. In this work, we focus on the resource synchronization protocols which are based on  (semi-) partitioned scheduling, detailed as follows:
	\begin{itemize}
		\item \textbf{Partitioned Schedule:} Each task is assigned on a dedicated processor, each 
		processor maintains its own ready queue and scheduler. Tasks are not allowed to
		migrate among processors, e.g., MPCP.
		\item \textbf{Semi-partitioned Schedule:} Unlike the pure partitioned schedule, semi-partitioned
		schedule allows tasks to migrate to other processors under certain conditions. For example,
		in DPCP and ROP-PCP, shared resources are assigned on processors, 
		the critical sections have to be executed on the corresponding processors, where
		may not be the same as the original partition of a task.
	\end{itemize}

	\noindent\textbf{Our Contribution in a nutshell:}
	We enhance the RTEMS with the aforementioned multiprocessor resource synchronization protocols and discuss how to revise the kernel with RTEMS Symmetric Multiprocessing (SMP) support. 
	\begin{itemize}
		\item To harden the open source development, we review the SMP support of RTEMS and point out the potential pitfalls during the implementation, so that the insights can be reused on any other platforms (see Section~\ref{sec:rtems-support}).
		\item We detail the development of three multiprocessor resource synchronization protocols, i.e., MPCP, DPCP, and FMLP, and their variants in RTEMS (see Section~\ref{sec:implementations}). 
		\item To study the impact of the implementation overheads, we deploy our implementations on a real platform with synthetic task sets (see Section~\ref{sec:overheads-evaluations}). 
		The measured overheads show 
		that our implementation overheads are comparable to the existed implementation of MrsP, in RTEMS, which illustrates
		the applicability of our implementations.	
	\end{itemize}
	The patches 	
	have been released
	under MIT license in~\cite{RTEMS-protocols} for RTEMS 4.12. Please note that this release branch was planned to be the latest release, but significant changes warranted to bump the major number from 4 to 5. To apply our patches to RTEMS 5, a certain adaption is additionally needed.
	
	\section{System Model}
	\label{sec:system-model}
	
	We consider a task set $\bf T$ consists of $n$ recurrent tasks to be scheduled on $M$ symmetric and 
	identical (homogeneous) processors. 
	All tasks can have multiple (non-nested) critical
	sections, each critical section accesses one of the 
	$Z$ shared resources, denoted as $s_z$ 
	Each task $\tau_i$ is described by a tuple
	\mbox{$(C_i,~\mu_i,~T_i,~D_i,~q_i)$}, where:
	\begin{itemize}
		\item $C_i$ is the worst-case execution time (WCET) of task $\tau_i$,
		i.e., $C_i > 0$.
		\item $~\mu_i$ is the set of resource(s) that $\tau_i$ requests.
		\item $T_i$ is the period of task $\tau_i$, i.e., $T_i > 0$.
		\item $D_i$ is the relative deadline of the task $\tau_i$. To fulfill its timing requirements
		a job of $\tau_i$ released at time $t$ must finish its execution before its
		absolute deadline $t+D_i$. We consider constrained-deadline task systems, 
		i.e., $D_i \leq T_i$ for
		every task $\tau_i \in \textbf{T}$.
		\item $q_i$ is the priority of task $\tau_i$.
		
	\end{itemize}
	
	
	\tikzstyle{decision} = [diamond,aspect=1.7, draw, fill=white, 
	text width=5em, text badly centered, node distance=2.15cm, inner sep=0pt]
	\tikzstyle{block} = [rectangle, draw, 
	text width=14em, text centered, rounded corners, minimum height=2.5em, node distance=1.5cm]
	\tikzstyle{line} = [draw,->]
	\tikzstyle{cloud} = [draw, ellipse,fill=red!20, minimum height=2em]
	
	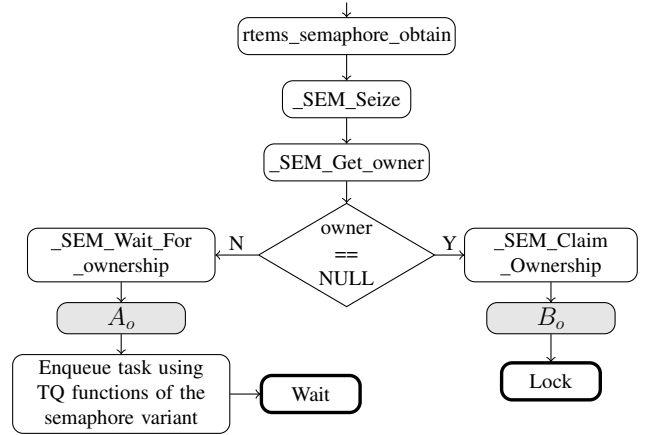
\begin{figure}[tp]
		\centering
		\begin{tikzpicture}[scale=0.56, transform shape]
			\tikzstyle{every node}=[font=\Large]
			\node [block] (init) {rtems\_semaphore\_obtain};
			\node [block, below of=init, text width=8em] (seize) {\_SEM\_Seize};
			\node [block, below of=seize, text width=10.5em] (getOwner) {\_SEM\_Get\_owner};
			\node [decision,  below of=getOwner, node distance = 2.2cm] (owner) {owner == NULL};
			
			\node [block,  left of=owner,node distance=5.35cm ,text width=12em] (wait) {\_SEM\_Wait\_For	\_ownership};
			\node [block,  below of=wait, text width = 8em, minimum height=2em, fill=gray!20] (semS) {\LARGE{\textbf{$A_o$}}};
			\node [block,  below of=semS,node distance=1.8cm, minimum height=4em ] (isNull) {Enqueue task using TQ functions of the semaphore variant};
			\node [block,line width=0.45mm,  right of=isNull, node distance=4.5cm, text width = 6em] (extract2) {Wait};
			\node [block,  right of=owner, node distance=4.85cm, text width = 11em] (notNull) {\_SEM\_Claim   \_Ownership};
			\node [block, ,  below of=notNull , text width = 8em, minimum height=2em, fill=gray!20] (semV) {\LARGE{\textbf{$B_o$}}};
			
			\node [block,line width=0.45mm,  below of=semV, text width = 6em] (extract) {Lock};
			
			\draw[<-] ([yshift=0pt]init.north)  -- ++(0em,1em);
			\path [line] (init) -- (seize);
			\path [line] (seize) -- (getOwner);
			\path [line] (getOwner) -- (owner);
			
			\path [line] (owner) -- node [midway,above] {Y}(notNull);
			\path [line] (owner) -- node [midway,above] {N} (wait);
			
			\path [line] (wait) -- (semS);
			\path [line] (semS) -- (isNull);
			\path [line] (notNull) -- (semV);
			\path [line] (isNull) -- (extract2);
			\path [line] (semV) -- (extract);
		\end{tikzpicture}
		\caption{Workflow of the lock directive. Block $A_o$ and $B_o$ are specified according to the adopted protocols.}
		\label{fig:flow_chart_lock}	
		\vspace{-0.15in}
	\end{figure}
	
	\section{Symmetric Multiprocessing Support in RTEMS}
	\label{sec:rtems-support}
	
	RTEMS allows users to implement new resource synchronization protocols by 
	strictly following the RTEMS API.
	To create a new semaphore, 
	\texttt{SEM\_Initialize} function
	is called to define the specified attributes for each resource synchronization protocol.
	Besides the creation of semaphore, which is defined by different protocols, some common 
	components that are similar for all the protocols, i.e., lock and unlock directives, configuration for applications, and migration
	mechanism, are introduced in this section.
	
	\subsection{Lock and Unlock Directives}
	\label{sec:lock-unlock}
	The workflow of the lock directive is shown in Fig.~\ref{fig:flow_chart_lock}.
	Once a task $\tau_i$ requests a shared resource, it will try to lock
	the corresponding semaphore.
	After selecting the right semaphore, denoted as \texttt{SEM},
	$\tau_i$ calls the
	\texttt{\_SEM\_Seize} function.
	Then, the ownership of the semaphore is checked by getting the owner of the Thread queue Control.
	If the semaphore is locked by another task, 
	$\tau_i$ has to wait for the owner to release the semaphore.
	The detailed operations in block $A_o$ are specified according to the design of different protocols.
	If there is no owner yet, $\tau_i$ is set as the owner of the semaphore, and 
	starts the execution of its critical section.
	The operations in block $B_o$ can be different depending on the specified design of protocols.

\begin{figure}[tp]
		\centering
		\begin{tikzpicture}[scale=0.56, transform shape]
			\tikzstyle{every node}=[font=\Large]
			\node [block, text width=15em] (init) {rtems\_semaphore	\_release};
			\node [block, below of=init, text width=10em] (seize) {\_SEM\_Surrender};
			\node [block, below of=seize, text width=12em] (getOwner)   {\_SEM\_Get\_owner};
			\node [decision,  below of=getOwner, node distance = 2.3cm, text width=10em, text height=0.1em] (owner) {owner\ != \\ executing};
			\node [block, line width=0.45mm,below of=owner, text width=8em, node        distance = 2.5cm] (release) {Not owner};
			\node [block,  right of=owner, node distance=5cm, text width = 8em,       fill=gray!20] (notNull) {\LARGE{\textbf{$A_r$}}};
			\node [block,line width=0.45mm,  below of=notNull, node distance=1.65cm,    text width = 6em] (unlock) {Unlock};
			
			\draw[<-] ([yshift=0pt]init.north)  -- ++(0em,1em);
			\path [line] (init) -- (seize);
			\path [line] (seize) -- (getOwner);
			\path [line] (getOwner) -- (owner);
			\path [line] (owner) -- node [midway,above,xshift = 0.25cm, yshift=-0.25cm] {Y} (release);
			\path [line] (owner) -- node [midway,above] {N} (notNull);
			\path [line] (notNull) -- (unlock);
		\end{tikzpicture}
		\caption{Workflow of the unlock directive. Block $A_r$ is specified according to the adopted protocols.}
		\label{fig:flow_chart_unlock}
		\vspace{-0.15in}
	\end{figure}
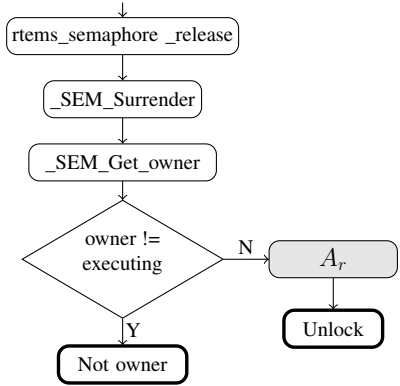

	The workflow of the unlock directive is shown in Fig.~\ref{fig:flow_chart_unlock}.
	It will be called when task $\tau_i$ has finished the execution of its critical section and 
	releases the lock of the semaphore.
	The unlock directive selects the right \texttt{\_SEM\_Surrender} function
	to check whether the $\tau_i$ is the current owner of the semaphore.
	If $\tau_i$ is not the owner, the semaphore cannot be unlocked.
	Otherwise, $\tau_i$ can unlock the semaphore by executing the commands
	in block $A_r$. 
	The main function in $A_r$ is to find the next owner for the semaphore if 
	(at least) one task that is waiting for the semaphore.
	If there is no waiting task, the owner will be set to \texttt{NULL} accordingly.
	The details of the functions in $A_r$ will be discussed 
	in the corresponding sections for different protocols.
	
	\subsection{Application Configuration}
	
\begin{figure}[ht]
		\centering
		\begin{tikzpicture}[scale=0.8, transform shape]
			\tikzstyle{every node}=[font=\large]
			\node[draw,black] (Processor) {Processor};
			\node[draw,black, node distance = 4.3cm, right of = Processor] (Scheduler) {Scheduler Instance};
			\node[draw,black,node distance = 4cm, right of = Scheduler] (Task) {Tasks};
			
			\path[draw,black,thick,dashed] (Processor) --node [midway,above] {Step 1} (Scheduler); 
			
			\path[draw,black,thick,dashed] (Scheduler) -- node [midway,above] {Step 2}(Task); 
			
		\end{tikzpicture}
		\caption{The steps to configure}
		\label{fig:rtems_config}
		\vspace{-0.1in}
	\end{figure}
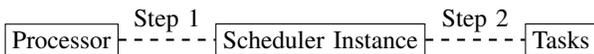

	In order to support semi-partitioned schedule in RTEMS, 
	the flow for configuration in Fig.~\ref{fig:rtems_config} has to be followed.
	Firstly, processors have to be bound to specific scheduler instances by using 
	macro \texttt{\_RTEMS\_SCHEDULER\_ASSIGN} supported in RTEMS by default.
	After that, each task is partitioned to a scheduler instance by using the \texttt{rtems\_task\_set\_scheduler} directive.
	Each task can only be executed on the processor of the corresponding scheduler instance.
	
	When a RTEMS application is configured with SMP support by following the work flow in Fig.~\ref{fig:rtems_config},
	some new functions have to be implemented.
	In Step 1, an initial task has to be defined, which is executed in the beginning of the RTEMS application.
	The binding of scheduler instances to processor is
	based of the guide in the official c-user guide. 
	The dedicated schedule algorithm for the scheduler instances has to be selected at first.
	In this paper, the \emph{Deterministic\ Priority\ SMP\ Scheduler} supported in RTEMS by default is selected for all the
	protocols, which is the same as Fixed-Priority (FP) scheduler in the literature.
	Please note that, the instances have to be defined for all the available processors in the system, in order to support
	the semi-partitioned schedule, i.e., tasks may migrate to other processors by changing their scheduler nodes,
	details can be found in next subsection. 
	
	\subsection{Migration Mechanism}
	\label{sec:migration-mechanism}
	
	The migration mechanism 
	by using arbitrary processor affinity in~\cite{DBLP:journals/rts/GujaratiCB15}
	is not supported
	in the current version of RTEMS. 
	Therefore, a new migration mechanism has to be applied for those distributed-based
	protocols, e.g., DPCP.
	In our implementation,
	the scheduler node is modified during the run time in order to
	realize the task migration.
	When a task needs to migrate to another processor, the scheduler node of the task 
	in its original scheduler instance is blocked, and the scheduler node of the task in its destination processor is unblocked.
	An additional function named \texttt{\_Scheduler\_Migrate\_To} is implemented in \texttt{schedulerimpl.h},
	which contains the task information block, the target processor,
	and the priority of the task in the target processor.
	In addition, in order to guarantee the correctness of the migration, \texttt{thread-dispatch} is disabled
	during the migration operation.
	
	Fig.~\ref{fig:migr} demonstrates an example of the implemented task migration.
	In Fig.~\ref{fig:migr}~(1),
	task $\tau_i$ has a scheduler node for every scheduler instance in the system.
	$\tau_i$ is currently executing on CPU\#0 with a priority of $7$ 
	by using scheduler node $S_0$, which is indicated by the the node with green background. 
	Other two nodes with grey background are blocked, since $\tau_i$ 
	has no access their respective scheduler instances, denoted as dashed line.
	In Fig.~\ref{fig:migr}~(2),
	task $\tau_i$ performs migration to CPU\#1.
	$\tau_i$ blocks itself on its original scheduler by using the block function of the scheduler instance on $S_0$.
	After that,
	it adds $S_1$ to the list of its active scheduler nodes and modifies the priority of $S_1$ accordingly.
	It unblocks $S_1$ by using the unblock function of the corresponding scheduler instance.
	Migrating back to the original processor works similarly, i.e.,
	Fig.~\ref{fig:migr}~(1) is restored by using the same unblock/block function of the scheduler instances.
	
	\tikzstyle{re} = [draw, rectangle, thick, black, fill = white,  align=center, text width = 9.5em]
	\tikzstyle{se} = [draw, rectangle, thick, black, fill = white,  align=center, text width = 7.5em] 
	\tikzstyle{line} = [draw,->]
	\begin{figure}[t]
		\centering
		\begin{tikzpicture}[scale=0.7, transform shape]
			
			\node[re, rounded corners = 4](thread){ {Task} $\tau_i$\\ {Executing on CPU\#0}};
			
			\node[se,below of=thread,fill=gray!20, dashed,yshift=-1em](s1)
			{$S_1${(255) [CPU\#1]} \\ {BLOCKED}};
			\node[se,left of=s1, xshift  = -6em, fill=green!20](s0)
			{$S_0${(7) [CPU\#0]}\\ {SCHEDULED}};
			\node[se,right of=s1,fill=gray!20,dashed, xshift=6em](s2)
			{$S_2${(255) [CPU\#2]}\\ {BLOCKED}};
			
			\path [line] (thread) -| (s0);
			\path [line,dashed] (thread) -- (s1);
			\path [line,dashed] (thread) -| (s2);
			
			\node[re, rounded corners = 4, below of =s1, yshift=-2em](thread2){ {Task} {$\tau_i$}\\ 
				{Migrated to CPU\#1}};
			
			\node[se,below of=thread2,fill=green!20, yshift=-1em](s12)
			{$S_1${(2) [CPU\#1]} \\ {SCHEDULED}};
			\node[se,left of=s12, xshift  = -6em, fill=gray!20](s02)
			{$S_0${(7) [CPU\#0]}\\ {BLOCKED}};
			\node[se,right of=s12,fill=gray!20,dashed, xshift=6em](s22)
			{$S_2${(255) [CPU\#2]}\\ {BLOCKED}};
			
			\path [line] (thread2) -| (s02);
			\path [line] (thread2) -- (s12);
			\path [line,dashed] (thread2) -| (s22);
			
			\node[left of = thread, xshift=-15em, yshift=0.5em]{(1)};
			\node[left of = thread2, xshift=-15em, yshift=0.5em]{(2)};
		\end{tikzpicture}
		\caption{Scheduler Node management: (1) Before migration, (2) After migration. Dashed blocks and lines represent that $\tau_i$ has no access to the respective scheduler instances, whereas green block is the currently used one.}
		\label{fig:migr}
		\vspace{-0.15in}
	\end{figure}
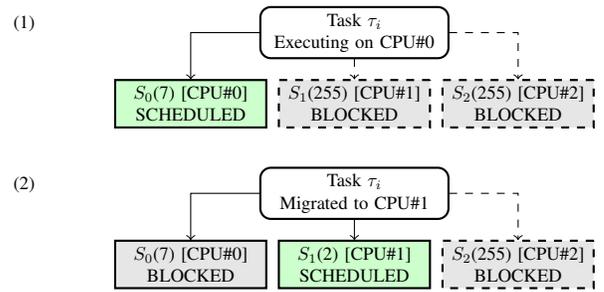
	
	\section{Multiprocessor Resource Synchronization}
	\label{sec:implementations}
	
	In this section, the implementation details of three protocols and corresponding
	variants are explained and discussed. Please note, 
	we only consider non-nested resource accesses in our implementation, i.e., only one 
	shared resource is requested during the execution of one critical section.
	
	\subsection{Multiprocessor Priority Ceiling Protocol}
	\label{sec:mpcp}
	The Multiprocessor Priority Ceiling Protocol (MPCP) is a typical protocol that is 
	based on a partitioned fixed priority (P-FP) scheduler. That is, each task
	has a pre-defined priority, 
	and the execution of a task is bound on a pre-defined processor, i.e., no migration is allowed.
	The main features of MPCP are:
	1) a task will suspend itself if the resource is not available.
	2) if a task is granted to access a shared resource, 
	the priority of the task will be boosted to the ceiling
	priority, which equals to the highest priority of these tasks that request that resource.
	
	The self-suspension feature is supported in RTEMS by default.
	In order to implement the ceiling priority boosting, one new semaphore structure is created.
	Besides these normal components, e.g., semaphore lock, wait queue, and current semaphore owner,
	one variable named \texttt{ceiling\_priority} is added. Please note that, in our implementation the ceiling priority is defined by users instead of being
		calculated by the system dynamically.
	The pseudo code provided in Algo.~\ref{alg:mpcp-semaphore} shows two main
	functions in our implementation, 
	which fits the lock and unlock directive in Section~\ref{sec:lock-unlock}.
	The details are as follows:
	Once a task $\tau_i$ requests a shared resource, 
	the ownership of the shared resource (semaphore) will be checked.
	If the owner of the requested shared resource is \texttt{NULL}, 
	$\tau_i$ becomes the owner, and the priority of $\tau_i$ is boosted to the ceiling priority on 
	the corresponding scheduler instance (operations in block $B_o$ in Fig.~\ref{fig:flow_chart_lock}). 
	Otherwise, $\tau_i$ will be added into a wait queue, which is sorted by tasks' original priorities, i.e., task with higher priority will get earlier position (operations in block $A_o$ in Fig.~\ref{fig:flow_chart_lock}).
	Once the task $\tau_i$ finishes the execution of critical section, 
	it will release the semaphore.
	The first task of the wait queue is checked, 
	i.e., the task with the highest priority in the wait queue.
	If there is no task in the wait queue, the semaphore owner will be set to \texttt{NULL}.
	Otherwise, the first task of the wait queue will be set as the semaphore owner (operations in block $A_r$ in Fig.~\ref{fig:flow_chart_unlock}).
	
	\begin{algorithm}[t]
		\caption{MPCP implementation}
		\label{alg:mpcp-semaphore}
		\begin{algorithmic}[1]
			\small
			\INPUT Task $\tau_{i}$, and 
			\texttt{ceiling\_priority} of related semaphore;
			\\~\\\textbf{Function} {mpcp\_lock}():
			\IF {\texttt{semaphore\_owner} is NULL}
			\STATE \texttt{semaphore\_owner} $\leftarrow$ $\tau_{i}$;
			\STATE ${\tau_i.priority}$ $\leftarrow$ \texttt{ceiling\_priority};
			\STATE $\tau_{i}$ starts the execution of its critical section;
			\ELSE 
			\STATE Add $\tau_{i}$ to the corresponding \texttt{wait\_queue};
			\ENDIF
			\\~\\\textbf{Function} {mpcp\_unlock}():
			\STATE $\tau_{i}$ releases the semaphore lock;	
			\STATE Next task $\tau_{next}$ $\leftarrow$ the head of the \texttt{wait\_queue};
			\IF { $\tau_{next}$ is NULL}
			\STATE \texttt{semaphore\_owner} $\leftarrow$ NULL;
			\ELSE 
			\STATE \texttt{semaphore\_owner} $\leftarrow$ $\tau_{next}$; 
			\STATE $\tau_{next}$ starts the execution of its critical section;
			\ENDIF	
		\end{algorithmic}
	
	\end{algorithm}
	
	\subsection{Distributed Priority Ceiling Protocol}
	\label{sec:dpcp}
	The Distributed Priority Ceiling Protocol (DPCP) is based on semi-partitioned fixed priority schedule.
	In DPCP, 
	tasks and shared resources are assigned on different processors separately, i.e., 
	these processors that are assigned for the execution of non-critical sections are called 
	application processors, and processors for the execution of critical sections are called synchronization processors.
	Once a task $\tau_i$ tries to access a shared resource, 
	it will migrate to the corresponding synchronization processor where the shared resource is assigned on, 
	before trying to lock the corresponding semaphore.
	Afterwards, 
	these tasks on the same synchronization processor operate follow the 
	uni-processor PCP, which been supported in RTEMS by default, i.e., 
	Immediate Ceiling Priority Protocol (ICPP). When a task $\tau_i$ finished its execution of critical section,
	it will migrate back to the original application processor to continue the execution of its non-critical section, 
	if it exists.
	
	Hence, the main challenge of the implementation of DPCP is to allow task migrations among processors.
	In RTEMS, task partitioning is realized by the scheduler node in the scheduler function, 
	i.e., scheduler node defines the original partition for each task before the execution, and 
	stays the same during the run time. Details have been explained in Section~\ref{sec:migration-mechanism}.

	\subsection{Flexible Multiprocessor Locking Protocol}
	\label{sec:fmlp}
	In Flexible Multiprocessor Locking Protocol (FMLP), requests of shared resources are divided into two groups, i.e., long and short, according to the
	length of the execution time of corresponding critical section.
	When the requested resource is not available, a task will suspend itself 
	if it is a long request, and a task will spin on the correspond processor if it is a short request.
	However, there is no conclusion regarding to how to divide requests to obtain a better schedulability.
	Therefore, we divided our implementation into FMLP-L which only supports long requests, and FMLP-S which only
	supports short requests.
	Please note, to simplify the implementation, 
	all the tasks in one task set all belong to either long group or short group, no mixed division
	of these two groups is allowed.
	
	In both FMLP-L and FMLP-S, the wait queue in the semaphore structure is in a FIFO order, rather than sorting by priorities like MPCP and DPCP.
	The operations in block $B_o$ in Fig.~\ref{fig:flow_chart_lock} are as follows:
	In FMLP-L, we maintain a ceiling priority dynamically for each resource, which equals to the highest priority 
	of these tasks that are currently 
	waiting for the resource, i.e., tasks in the corresponding wait queue.
	The priority of the semaphore owner will be boosted to the ceiling priority if the original priority is lower than the ceiling priority,
	when it starts the execution of its critical section.
	In FMLP-S, the owner of the semaphore gets priority boosted to the highest possible priority 
	in the system, so that the execution of its critical section is the non-preemptive.
	The operations in block $A_o$ in Fig.~\ref{fig:flow_chart_lock} are the same for both FMLP-L and FMLP-S, 
	i.e., add task $\tau_i$ in the end of the corresponding wait queue.
	The unlock operations in block $A_r$ in Fig.~\ref{fig:flow_chart_unlock} are also the same, i.e., 
	try to find the next owner for the semaphore by
	checking the first task in the wait queue, if it exists.
	
	Additionally, we implemented a distributed version of FMLP, denoted as DFLP, 
	where all the requests
	are treated as long requests.
	The main difference between FMLP and DFLP is when a task requests a shared resource, 
	it will migrate to the 
	corresponding synchronization processor, which is similar to DPCP. 
	The mechanism how we implement the migration has been
	explained in Section~\ref{sec:migration-mechanism}.
	After the migration, critical sections are executed by following the FMLP-L on the corresponding
	synchronization processor(s).
	
	\section{Evaluation and Discussion}
	\label{sec:overheads-evaluations}
	
	In this section, we introduce the setup of experiments for overheads evaluation at first.
	Afterwards, the measured overheads are reported and analyzed. At the end, we discuss the need of formal verification over the implementation generally.
	
	\subsection{Experimental Setup}
	
	We evaluated the overheads of our implementations on the
	following platform:
	a NXP QorIQ T4240 RDB reference design board, which is the same as used in~\cite{DBLP:conf/adaEurope/CatellaniBHM15}.  
	It has 6 GB DDR3 memory with 1866 MT/s data rate, 
	128 MB NOR flash(16-bit), and
	2 GB SLC NAND flash.
	The processor T4240 contains 24-virtual-core (12 physical cores) with the PowerPC Architecture, and is running on 1.67 GHz. 
	
	To measure the overheads of our implemented protocols,
	timestamps are added before and after the function of our implementations.
	The obtain and release functions of the semaphore are measured, denoted as lock and unlock respectively.
	We consider a multi-processor system consists of four processors, 
	i.e., $M = 4$, including three application processors and 
	one synchronization processor.
	The total number of tasks $n=15$, and the number of available shared resources $Z=3$, 
	i.e., $\mu_i \in \{s_1, s_2, s_3\}$.
	On each application processor,
	there are five tasks with five different priority levels, i.e.,
	$q_i \in$ \{High (H), Medium-High (MH), Medium (M), Medium-Low (ML) and Low (L)\}.
	Each task requests one of these three shared resources.
	Details can be found in Table~\ref{tab:cpu_allocation}.

	\begin{table}[t]
		\centering
				\caption{\normalsize Processor allocation of the test application.}
		\begin{tabular}{|p{15.5mm}|p{15.5mm}|p{15.5mm}|p{21.5mm}|}
			\hline
			CPU\#0 \newline Application 
			& CPU\#1 \newline Application & CPU\#2\newline Application
			& CPU\#3 \newline Synchronization \\
			\hline
			L ($s_1$) & L ($s_2$) & L ($s_3$) & -\\
			\hline
			ML ($s_2$) & ML ($s_3$) & ML ($s_1$) & -\\
			\hline
			M ($s_3$) & M ($s_1$) & M ($s_2$) & -\\
			\hline
			MH ($s_2$) & MH ($s_3$) & MH ($s_1$) & -\\
			\hline
			H ($s_3$) & H ($s_1$) & H ($s_2$) & -\\
			\hline
			
		\end{tabular}
		\label{tab:cpu_allocation}
		\vspace{-0.15in}
	\end{table}

	%

	\begin{figure*}[t]
		\centering
		\includegraphics[width=0.95\linewidth]{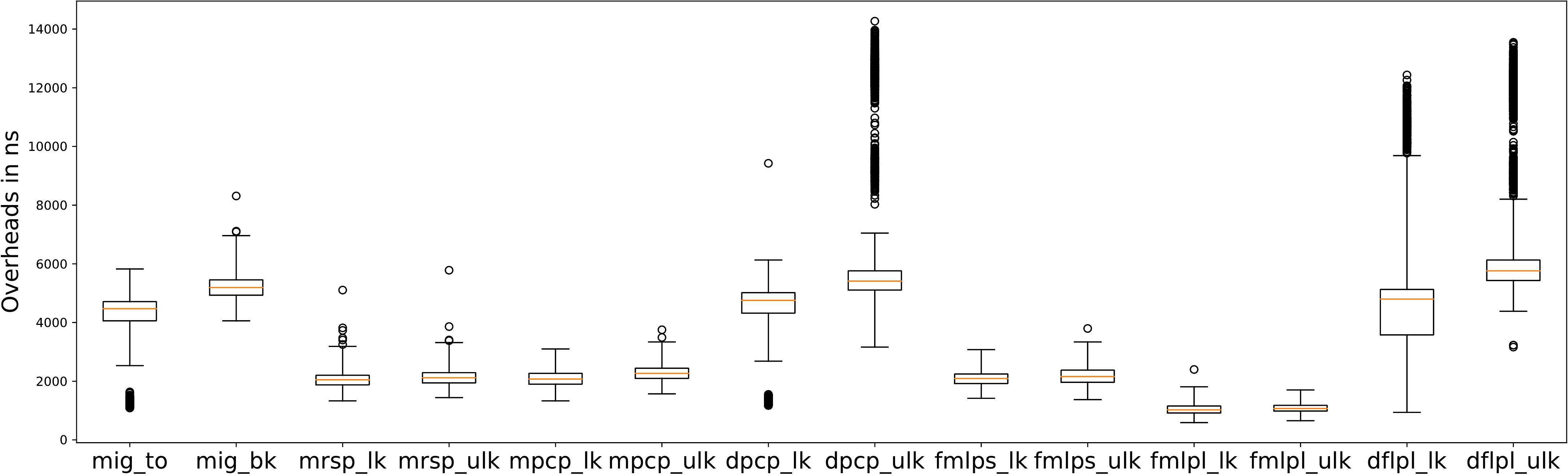}
		\caption{Overheads of protocols in RTEMS (lock operation is ended by \texttt{\_lk} and unlock operation is ended by \texttt{\_ulkl})). The measurement of migrating a task to the synchronization processor (denoted as \texttt{mig\_to}) and back to the application processor (denoted as \texttt{mig\_bk}).}
		\label{fig:overheads}
		\vspace{-0.1in}
	\end{figure*}
	
	\subsection{Overheads Evaluation}
	
	The overheads for different protocols are reported in Figure~\ref{fig:overheads},
	based on more than 9,000 instances of lock and unlock operations.
	These distributed-based protocols, i.e., DPCP and DFLP have higher overheads than others, due to the task migrations. 
	DFLP has the highest average overheads, since it also maintains the dynamic ceiling priority update.
	MrsP also has relative high overheads, since it has the help mechanism requiring task migration (however, help mechanism may not be activated all the time). Our results related to MrsP are similar as reported in~\cite{DBLP:conf/adaEurope/CatellaniBHM15}, i.e., 
	5376 ns for lock and 5514 ns for unlock on average.
	FMLP-L has the lowest overheads, due to the simplest mechanism. Overall, the overheads for all the protocols are relatively low and acceptable.
	For distributed-based protocols, we can observe that there are quite a few outliers. In fact, a similar observation has been reported in~\cite{shi-OSPERT17}. One reason could be that the behavior of cache memories kicks in to introduced operation overheads, but we have no sufficient data to pinpoint the exact cause here.
	
	
	The migration overheads are measured separately, and reported in the left side of Figure~\ref{fig:overheads}. The results show that the overheads of task migration
	are significant, which might substantially affect those distributed-base protocols, i.e., DPCP and DFLP.
	Interestingly, we also notice that the overhead of a task to migrate to the synchronization processor is faster than migrating back to the application processor. The reason is that, normally there are more tasks running on the application processors than synchronization processors, which causes a task has to wait for longer time to obtain
	the scheduler instance lock on average.
	That is why the unlock overheads of DPCP and DFLP are higher than the lock overheads.
	
	Although our evaluated overheads on RTEMS are similar to these protocols that are implemented on \litmus~\cite{DBLP:conf/adaEurope/CatellaniBHM15, Shi-Dependency-RTAS2019}, implementations of protocols on RTEMS and \litmus are not directly comparable due to the difference of purposes and architectures in two operating systems. Please note that RTEMS is a self-contained RTOS for real-world applications, whilst \litmus is a Linux-based testbed, which is mainly used for functional validation. 
	It might be interesting to investigate which protocol is preferable on which operating systems, but it is considered out of scope here.

	\subsection{Validation and Formal Verification}
	
	To validate the correctness of our implementation, at first we test over the official coverage tests provided by RTEMS, i.e., the SMP test suites (\url{https://github.com/RTEMS/rtems/tree/master/testsuites/smptests}) especially, on the PowerPC device and also the QEMU emulator for ARM RealView Platform \texttt{realview-pbx-a9}, and conclude that the SMP related peripheries in RTEMS are not affected at all. Moreover, we further design several dedicated corner cases for each protocol and ensure that the designated tasks execute as the expected behaviors, which are 
	treated
	as the additional coverage test for the future integration.
	
	We note that such case-based validation may not be sufficient, since it is not possible to test over every case exhaustively. One possible way is to adopt software model checkers as proposed in ~\cite{DBLP:conf/icfem/GadiaAB16} to detect potential data races and deadlocks in the implementation of PIP with nested locks in RTEMS. However, such searching approaches may not scale well for multiprocessor protocols unless an effective pruning strategy can be found beforehand. How to validate or formally verify an existing implementation of synchronization protocols is still an unsolved problem but out of the scope.
	
	\section{Conclusion}
	\label{sec:conclusions}
	Over the decades, quite a few number of resource synchronization protocols have been extensively studied for uni-processor and especially multiprocessor real-time systems. In this work, we reviewed the SMP support in one popular real-time operating system RTEMS and detailed how we develop three state-of-the-art multiprocessor resource synchronization protocols, i.e., MPCP, DPCP, and FMLP, and their variants. With extensive synthetic experiments, the measured results showed that our implementations are comparable to MrsP, which is officially supported in RTEMS. Considering the real system overhead, the performance of resource synchronization protocols might be clarified and decidable by system designers. 
	
Although several dedicated tests are provided to verify the correctness of the implementation, formal model checking is still desirable to prevent the system from potential deadlock, data races, and priority inversions. In the future work, we plan to explore on nested resource synchronization and support the arbitrary processor affinity in RTEMS to improve the generality and the efficiency. An ongoing effort is also provided to support for the latest version of RTEMS.
	
	
	

		\section*{Acknowledgement}
		{This paper is supported by DFG, as part of the
			Collaborative Research Center SFB876, subproject A1 and A3	{(http://sfb876.tu-dortmund.de/)}.}

	\bibliographystyle{abbrv} \bibliography{real-time}

\end{document}